\documentclass[varenna]{cimento}
\usepackage{latexsym}
\usepackage{amsmath,amssymb}
\usepackage[retainorgcmds]{IEEEtrantools}
\usepackage[T1]{fontenc}
\usepackage{graphicx}
\usepackage{sidecap}

\title{Single-top production at hadron colliders}
\author{P.~Motylinski}
\institute{Physikalisches Institut, Albert-Ludwigs-Universit\"at Freiburg, Hermann-Herder-Stra\ss e 3, D-79104 Freiburg i.Br., Germany}
\PACSes{\PACSit{14.65.Ha}{Top quarks}
\PACSit{12.15.-y}{Electroweak interactions}}

\begin{document}
\maketitle
\begin{abstract}
We review the recent theoretical progress in single-top physics at hadron colliders. Apart from single-top production within the Standard Model we treat certain aspects of single-top production in beyond Standard Model scenarios.
\end{abstract}

\section{Introduction}
\begin{center}
  \begin{figure}
    \label{fig:st_lo}
    \begin{center}
      \includegraphics[width=0.8\textwidth]{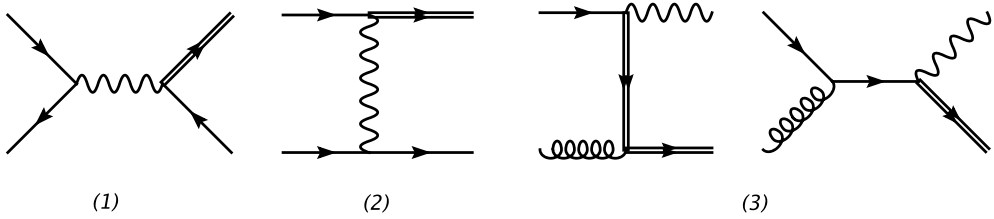}     
      \caption{Leading order diagrams for single top production in the Standard Model. {\bf \emph{(1)}} and {\bf \emph{(2)}} are the $s$-- and $t$--channel diagrams, respectively. {\bf \emph{(3)}} are the LO diagrams for $Wt$-associated production.}
    \end{center}
  \end{figure}
\end{center}

Single-top production in the Standard Model happens through electroweak (EW) interactions. The cross section for single-top productions is sizable, being roughly one third that of $t\bar{t}$--production at the LHC. \\
Studying single-top production gives a unique possibility to study the physics of a \emph{single} quark as well as fundamental properties of the Standard Model such as the chirality of the $Wt$--coupling and, by measuring $V_{tb}$, the unitarity of the CKM matrix. Due to strong correlations between the top quark and its decay products, single-top production is sensitive to right-handedness in both production and decay vertices. Single-top production is also expected to play a crucial role in determining the $b$--quark content in protons and anti-protons.\\
Within the Standard Model there are overall three production channels, shown in fig.(\ref{fig:st_lo}). Of those the $t$--channel (diagram 2) is by far the largest at both the Tevatron and the LHC. The $s$--channel contribution (diagram 1) is significant at the Tevatron but rather small at the LHC. For the associated $Wt$--production (diagrams 3) the situation is reversed, i.e. it is negligible at the Tevatron while being significant at the LHC, due to the large gluon content of the protons at LHC energies.\\

\section{Single top production at NLO}
\label{sec:MCNLO}
\emph{Next-to-leading order} (NLO) corrections to the total cross section of $s$-- and $t$--channel single-top productions have been known for some time already~\cite{Bordes:1994ki,Smith:1996ij,Stelzer:1997ns}. They were followed by NLO corrections to the fully differential cross sections~\cite{Harris:2002md,Sullivan:2004ie}. More recently single-top production has been included in Monte Carlo (MC) event generators which are based on next-to-leading order (NLO) cross sections. There are two main frameworks in this context: MC@NLO~\cite{Frixione:2002ik} and POWHEG~\cite{Frixione:2007vw} which are both extensively used by experimental collaborations. In both, all three production channels shown in fig.(\ref{fig:st_lo}) are included, including the decay of the top quark and angular correlations~\cite{Frixione:2005vw,Frixione:2007zp,Frixione:2008yi,Alioli:2009je,Re:2010bp}. We shall return to a more detailed description of the $Wt$--associated production and $Ht$--associated production.\\
Both MC@NLO and POWHEG show overall excellent agreement for a large range of observables. Here, we refer to~\cite{Frixione:2005vw,Frixione:2007zp,Frixione:2008yi,Alioli:2009je,Re:2010bp} for details and comparisons. 

\subsection{$Wt$--associated production}
\begin{center}
  \begin{figure}
    \begin{center}
    \label{fig:wt_int}
    \includegraphics[width=0.6\textwidth]{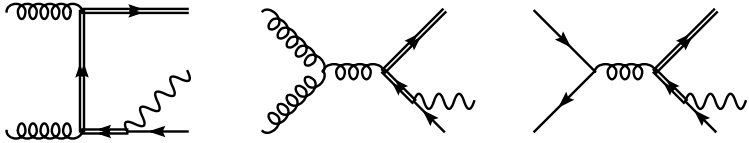}
    \caption{NLO $W^-t$--associated production interference with LO $t\bar{t}$--production (in the case of an on-shell $\bar{t}$--quark in the latter).}
       \end{center}
  \end{figure}
\end{center}

The third production channel shown in fig.(\ref{fig:st_lo}) is the $W^-t$--associated production channel. As can be seen there are overall two topologies contributing to this production mode at LO, and in both cases there is one gluon in the initial state. At the Tevatron the $W^-t$--cross section is negligible but at the LHC, due to the high gluon content of the colliding hadrons at those energies, $W^-t$--production is the second largest of the three channels in fig.(\ref{fig:st_lo}) with $\sigma_{W^-t}(\sqrt{s}=14 \textrm{TeV})\sim$60 pb.\\
From a theoretical point of view $W^-t$ associated production is rather subtle and hence deserves a bit more attention here. At NLO $W^-t$--associated productions interferes with LO $t\bar{t}$--production in the limit where the $\bar{t}$--quark becomes on-shell and decays into $W^-\bar{b}$. The interfering diagrams are shown in fig.(\ref{fig:wt_int}). 
In order to define $W^-t$--production in its own right it is necessary to provide a method that makes the distinction from $t\bar{t}$--possible. This is a well-known issue and has been described in various papers~\cite{Tait:1999cf,Belyaev:1998dn,Kersevan:2006fq,Zhu:2002uj,Campbell:2005bb,Cao:2008af}.\\ 
Theoretically, a consistent procedure would be to consider all processes having the final states $W^+ W^+ b \bar{b}$ or $W^+ W^+ b$ and then halt the perturbative expansion at $\mathcal{O}(\alpha_S^2 \alpha)$. This, however, would be at the expense of the theoretical accuracy of $t\bar{t}$--production. Therefore a different approach was chosen within both the MC@NLO and POWHEG frameworks and we briefly sketch the ideas here.\\
As mentioned the aim is, ultimately, to be able to generate two physically well-defined samples for both $W^-t$-- and $t\bar{t}$--production without facing the problem of interference. Furthermore, both should be defined in the context of a parton shower MC. The approach chosen in both MC@NLO and POWHEG is to consider two fundamentally different definitions of $W^-t$--production and to compare them~\cite{Frixione:2008yi,Re:2010bp}. Identifying the interference term as $\mathcal{I}$, the doubly resonant contribution as $\mathcal{D}$ and the singly resonant contribution as $\mathcal{S}$, the first approach, called \emph{Diagram Removal} (DR), consists of removing the contributions corresponding to terms $\mathcal{I}$ and $\mathcal{D}$.
While this obviously removes the overlap with $t\bar{t}$--production it formally breaks QCD gauge invariance. The issues connected to this are clarified in great detail in~\cite{Frixione:2008yi}.\\
In order to assess the impact of the shortcomings of DR a second approach has been taken, called \emph{Diagram Subtraction} (DS). In this approach a counter-term is constructed such that the doubly resonant parts are subtracted effectively when $t\bar{t}$ becomes resonant. 
This approach is gauge invariant and the subtraction is performed at the level of the amplitude squared. Considering the expressions for both DR and DS we can say something about the size of the interference term and hence something about the validity and possible shortcomings of both approaches. In~\cite{Frixione:2008yi} the two approaches were compared and it was shown that there is a very good agreement. This way $W^-t$--associated production has been given a meaningful definition in the context of NLO MC frameworks, allowing for detailed studies of this important channel.
\label{sec:wt}

\subsection{Interference effects between production and decay}
\label{sec:st_SCET}
In the context of single-top production and decay most approaches rely on the \emph{Narrow-Width Approximation} (NWA). In NWA the top is produced on-shell and, then, subsequently decays. NWA has obvious advantages since it simplifies the calculation significantly. It is possible to include NLO corrections for production and decay but these are only limited to the so-called \emph{factorizable} corrections.\\
In~\cite{Falgari:2010sf,Falgari:2011qa} \emph{Soft-Collinear Effective Theory} (SCET) was used to calculate NLO corrections for both production and decay, including both factorizable and non-factorizable corrections, this way assessing the impact of the interference between production and decay of the top. It was found that the corrections stemming from the non-factorizable corrections are, in general, small, thus rendering the NWA approach a valid approximation for most purposes.

\section{Single-top production beyond NLO}

\subsection{Single-top $t$--channel + 1 jet}
\label{sec:st_1jet}
\begin{center}
  \begin{figure}
    \label{fig:st_1jet}
    \begin{center}
      \includegraphics[width=0.4\textwidth]{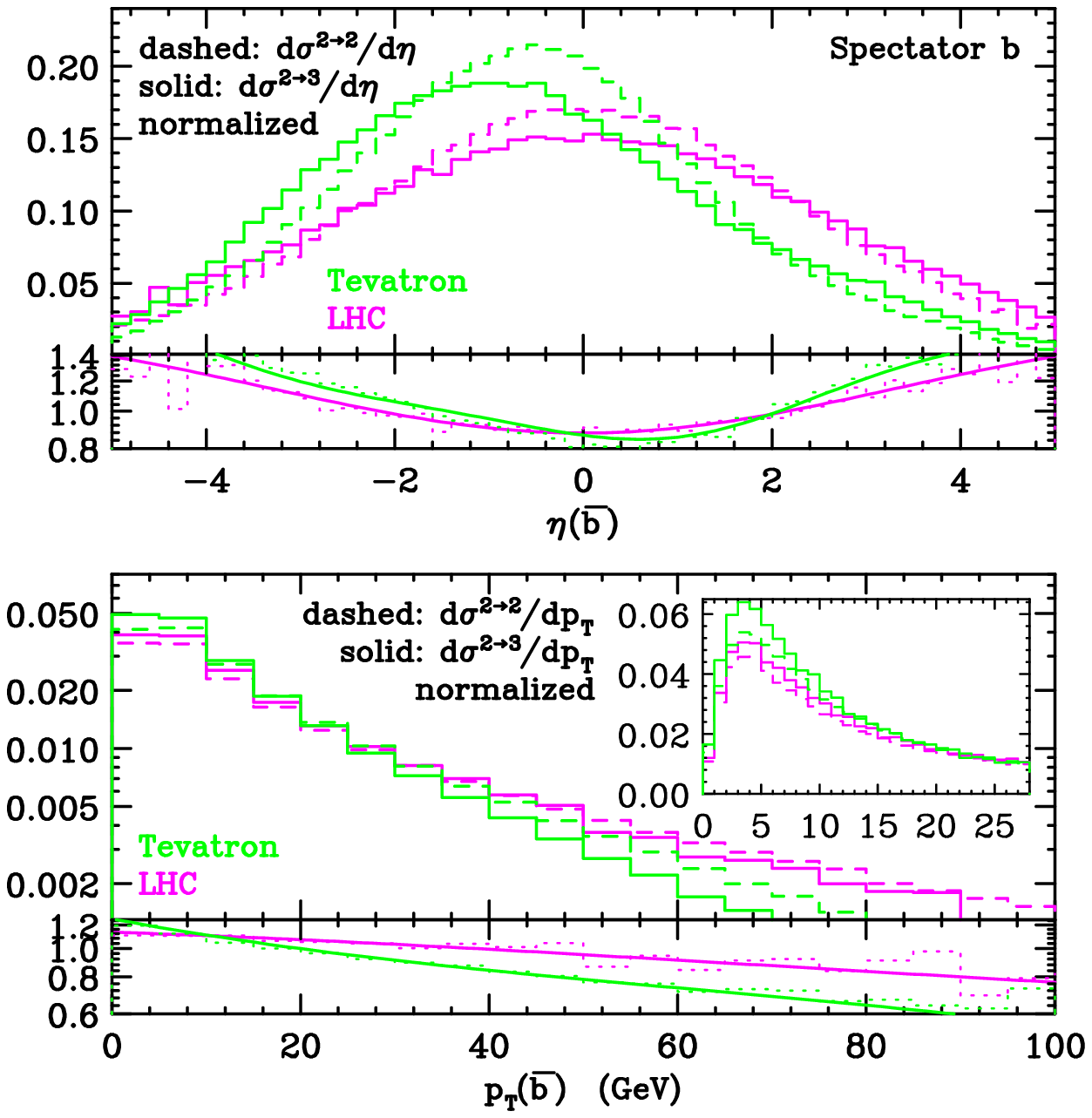}     
      \includegraphics[width=0.4\textwidth]{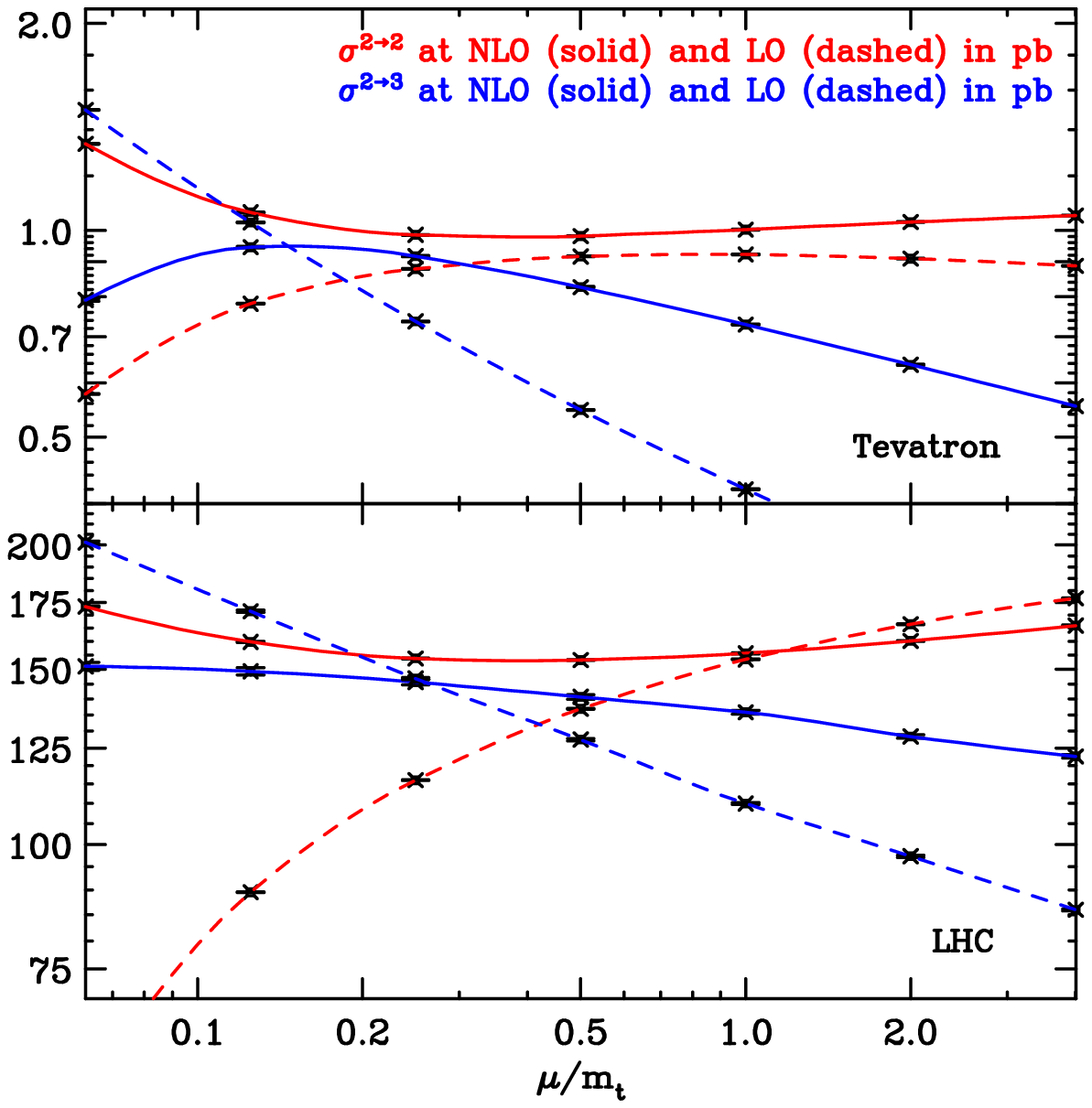}
      \caption{Comparing the 4F and 5F schemes~\cite{Campbell:2009ss}. \emph{Left}: rapidity and $p_T$--distributions for the spectator $b$--quark. \emph{Right}: scale dependence.}
    \end{center}
  \end{figure}
\end{center}

Current versions of both MC@NLO and POWHEG rely on the five flavor scheme (5F). In this approach the LO $t$--channel diagram is that of a $2\rightarrow2$ process of type $qb\rightarrow q' t$ or $\bar{q}b \rightarrow \bar{q}' t$. The $b$--quark is then extracted from a PDF library. At NLO the $b$--quark can also originate from the splitting of a gluon thus contributing to the $2\rightarrow 3$ processes  $qg\rightarrow q' t \bar{b}$ or $\bar{q} g \rightarrow \bar{q}' t \bar{b}$. The 5F scheme has obvious advantages. The calculation is considerably simplified, the possible collinear divergences of the $g\rightarrow b \bar{b}$ splitting are resummed in the PDF and issues connected to the presence of a $b$-jet is dealt with at NLO. \\
Alternatively it is possible to view the $2\rightarrow 3$ as being of LO and then calculate the NLO corrections~\cite{Campbell:2009ss,Campbell:2009gj}.  This means that the $b$--quarks, which are allowed to be massive, are not part in the QCD evolution of the PDF's. In other words, the LO calculation is carried out in the four flavor scheme (4F). The two schemes are equivalent if an all-order calculation would be performed but at fixed order a comparison allows for a closer inspection of the role played by the final state $b$--quark coming from gluon splitting (the spectator $b$--quark). \\
In~\cite{Campbell:2009ss} a comparison between the two schemes was carried out for both the Tevatron ($p\bar{p} @ \sqrt{s}=1.96$TeV) and the LHC ($pp @ \sqrt{s}=14$TeV) and using $m_t=172$GeV, $m_b=4.7$GeV and the CTEQ6.6 PDF set~\cite{Nadolsky:2008zw}. In fig.(\ref{fig:st_1jet}) we see the rapidity and $p_T$ distributions for the spectator $b$--quark and the scale dependence for both the 4F and 5F schemes. For the $b$--quark we see that there are small yet visible differences between the two schemes. Furthermore we also see the increased sensitivity to the scale choice in the 4F scheme, which stems from the additional, explicit dependence on $\alpha_S$.

\subsection{Approximate NNLO}
\label{sec:approxNNLO}
While fixed-order perturbative calculations order have improved a large range of predictions there are regions of phase space that remain poorly described. An example is soft- and collinear radiation near threshold. Typically logarithms of the type $\ln^k(s_4/m^2)/s_4$, where $s_4=s+t+u-m^2$, tend to increase considerably, making effects from such infrared emissions large, and it is thus necessary to resum these contributions. \\
Recently, progress has been reported in providing results accurate to \emph{next-to-next-to leading logarithm} (NNLL)~\cite{Kidonakis:2010tc,Kidonakis:2011wy,Kidonakis:2010ux}. After a Mellin transform the resummed cross section takes on the general form
\begin{eqnarray}
  \label{eq:NNLO_res}
  {\hat{\sigma}}^{res}(N) &=&   
  \exp\left[ \sum_{i=1,2} E(N_i)\right] \exp\left[ {E'}(N')\right] \exp \left[\sum_{i=1,2} 2 \int_{\mu_F}^{\sqrt{s}} \frac{d\mu}{\mu} \gamma_{q/q}\left({\tilde N}_i, \alpha_s(\mu)\right)\right] \nonumber \\
  && \hspace{-10mm} \times {\rm Tr} \Bigg \{H\left(\alpha_s(\sqrt{s})\right) \exp \left[\int_{\sqrt{s}}^{{\sqrt{s}}/{\tilde N'}} \frac{d\mu}{\mu} \;\Gamma_S^{\dagger} \left(\alpha_s(\mu)\right)\right] \times S\left(\alpha_s(\sqrt{s}/{\tilde N'}) \right) \exp \left[\int_{\sqrt{s}}^{{\sqrt{s}}/{\tilde N'}} \frac{d\mu}{\mu} \Gamma_S \left(\alpha_s(\mu) \right) \right] \Bigg \} 
\end{eqnarray}
Apart from the process independent exponents resumming soft and collinear gluon radiation ($E(N_{1,2})$, ${E'}(N')$), the hard scattering function $H$ and the soft function $S$ describing soft non-collinear emissions, and the parton density anomalous dimension $\gamma_{q/q}$ the calculation involves calculating the soft anomalous dimension $\Gamma_S$ to two loops. The NNLL cross section is subsequently expanded to $\mathcal{O}(\alpha_S^2)$ in order to obtain the approximate \emph{next-to-next-to leading order} (NNLO) result.\\
While the approximate NNLO corrections in the case of the $t$--channel remain moderate, they become large for both the $s$--channel and the associated $Wt$--production. The impact of the corrections is shown in table~\ref{tab:appNNLO}.
\begin{table}
  \caption{Single-top cross sections incl. approximate NNLO corrections~\cite{Kidonakis:2010tc,Kidonakis:2011wy,Kidonakis:2010ux}. The corrections (in parentheses) are wrt. NLO corrections. The results are for $m_t=173$GeV, using the MSTW2008NNLO PDF set~\cite{Martin:2009iq}. Scale and PDF uncertainties, repsectively, are also shown.}
  \label{tab:appNNLO}
  \begin{tabular}{lccc}
      \textbf{$\sigma$} (pb) & Tevatron & LHC (7 TeV) & LHC (14 TeV)  \\
      \hline\\
      \textbf{$s$--channel} & $0.523^{+0.001+0.03}_{-0.005-0.028}$ \textcolor{blue}{($+15\%)$} & $3.17 \pm 0.06^{+0.13}_{-0.10}$ \textcolor{blue}{($+13\%$)} & $7.93 \pm 0.14^{+0.31}_{-0.28}$ \textcolor{blue}{($+13\%$)} \\
      \textbf{$t$--channel} & $1.04^{+0.00}_{-0.02}\pm 0.06$ \textcolor{blue}{($+4\%$)}& $41.7^{+1.6}_{-0.2}\pm 0.8$\textcolor{blue}{($-1\%$)} & $151^{+4}_{-1}\pm 3$ \textcolor{blue}{($-3\%$)}\\
      \textbf{$Wt$--prod.} & omitted & $7.8 \pm 0.2^{+0.5}_{-0.6}$ \textcolor{blue}{($+8\%$)} & $41.8 \pm 1.0^{+1.5}_{-2.4}$ \textcolor{blue}{($+8\%$)} \\
      \hline
  \end{tabular}
\end{table}

\section{Beyond the Standard Model}
\label{sec:beyond_sm}

\subsection{Charged Higgs production in association with a top quark}
\label{sec:Ht_prod}

From a technical point of view the most obvious extension to the Standard Model is to consider the associated production of a top quark with a charged Higgs boson, i.e. $H^-t$--production. Charged Higgs bosons occur naturally in a range of models, including \emph{Two Higgs-Doublet Models} (2HDM) and in the \emph{Minimal Super-symmetric Model} (MSSM).\\
As indicated, $H^-t$--productions is technically very similar to $W^-t$--production. However, since charged Higgs bosons have not yet been observed, it is necessary to distinguish between two case:
\begin{enumerate}
\item $m_H > m_t$: $H^-t$ is produced directly in association with a top. The mass condition forces the anti-top to be strictly off-shell.
\item $m_H < m_t$: $H^-t$ interferes with $t\bar{t} \rightarrow t H^- b$.  
\end{enumerate}
In the latter case one faces the exact same technical challenges as in $W^-t$--production. Here it is sufficient to point out that it has been implemented in MC@NLO using the same definitions (DR and DS)~\cite{Weydert:2009vr}. More recently, approximate NNLO results for the $H^-t$--cross section have also been calculated~\cite{Kidonakis:2010ux}. \\
While being technically easier, the first case, i.e. for $m_H>m_t$, presents other challenges, connected to signal extraction. The suggested procedure in previous studies requires the presence of a second $b$--jet to be used as a handle in $H^-t$--event selection. However the investigation carried out in~\cite{Weydert:2009vr} showed that given that one hard $b$--jet has been observed the probability that a second $b$--jet is present is only $\simeq$35\% for leptonic top decays and $\simeq$12\% for hadronic decays. Using the hardness of $b$--jets thus does not appear as a strong handle.

\subsection{Anomalous couplings}
\label{sec:anom_coupl}

The Standard Model predicts the $Wt$--coupling to be purely left-handed. However, it still remains to be determined whether the coupling also has a right-handed component. By conducting detailed studies of single-top production and decay it will be possible to learn more about the nature of the $Wtb$--coupling.\\
Theoretical studies are often approached by considering a model-independent parametrization of the following kind (for $t(p) \rightarrow b(k) + W^+(q)$ and $b(k) \rightarrow t(p) + W^-(q)$, respectively):
\begin{eqnarray}
  \label{eq:anom1}
  V_{t\rightarrow bW^+} =& \frac{-g}{\sqrt{2}}V_{tb} \Big [ \gamma^{\mu}(f_{1L}P_L+f_{1R}P_R) - \frac{i\sigma^{\mu \nu}}{m_W}(p-k)_{\nu} (f_{2L}P_L + f_{2R}P_R) \Big ]\\
  V_{b\rightarrow tW^-} =& \frac{-g}{\sqrt{2}}V_{tb} \Big [ \gamma^{\mu}(f^*_{1L}P_L+f^*_{1R}P_R) - \frac{i\sigma^{\mu \nu}}{m_W}(p-k)_{\nu} (f^*_{2L}P_L + f^*_{2R}P_R) \Big ]
\end{eqnarray}
where $f_{1L}=1$, $f_{1R}=f_{2L}=f_{2R}=0$ in the Standard Model. A study carried out recently for $Wt$--productions found that the azimuthal and energy distributions for the charged lepton (from the decaying top) and the azimuthal distribution of $b$--quark are sensitive to $\textrm{Re}(f_{2R})$ and $\textrm{Im}(f_{2R})$ in the $Wtb$--coupling~\cite{Rindani:2011pk}. \\
Another approach is to consider the angle between the charged lepton in the $W$ rest frame and the $W$--momentum in the top rest frame, $\theta_l^*$. Using this observable the partial width of the top quark can be written as:
\begin{eqnarray}
  \frac{1}{\Gamma} \frac{d\Gamma}{d \cos \theta_l^*} =& \frac{3}{8} (1+\cos \theta_l^*)^2 F_R + \frac{3}{8} (1-\cos \theta_l^*)^2 F_L + \frac{3}{4} \sin^2 \theta_l^* F_0
\end{eqnarray}
where $F_i=\Gamma_i/\Gamma$ is the normalized partial width for top decay to the three $W$ helicity states. By measuring the $F_i$'s it is then possible to constrain the $Wtb$--coupling. Furthermore, it is possible to measure $\rho_{R,L}=\frac{F_{R,L}}{F_0}$.
Recently this approach has been refined~\cite{AguilarSaavedra:2010nx}. The authors showed that it is possible to define an asymmetry, $A^N,$ that is sensitive to variations in $\textrm{Im} f_R$ and with $A^N=0$ in the SM.\\

\subsection{Flavor Changing Neutral Currents (FCNC)}
\label{sec:fcnc}

New physics in the top sector might manifest itself through top production happening via FCNC. In~\cite{Gao:2009rf} NLO QCD corrections were calculated for single-top production happening via model-independent $tqg$--couplings at hadron colliders. It was shown that for $tcg$--coupling the corrections can enhance the total cross section by 60\% at the Tevatron and 30\% at the LHC, while the corresponding numbers for the $tug$--coupling are 50\% and 20\%.\\
By studying top production in association with jets it is possible to learn about the nature of the couplings of the top quark. In~\cite{Plehn:2009it} direct top production was compared to two SM single-top production channels (single-top $s$-- and $t$--channel). In particular angular correlations between the leptons, stemming from the top decay, and jets were investigated and it was shown that direct top production often shows different correlations from its SM brethren.\\
More recently production of top quarks together with missing energy, so-called \emph{monotops}, was investigated~\cite{Andrea:2011ws}. Production of monotops cannot happen at tree-level within the SM and their discovery would therefore imply new physics. For a more detailed discussion of the discovery potential and constraints we here refer to~\cite{Andrea:2011ws}.\\
We refer to~\cite{Bernreuther:2008ju} for further discussions on possible FCNC scenarios and for further references.\\

\acknowledgments
The author wishes to thank the organizers for the invitation to the TOP2011 Workshop in Sant Feliu de Gu\'ixols.

\end{document}